\documentclass[aps,prl,nobibnotes,twocolumn,superscriptaddress,bibliography]{revtex4-2}

\usepackage{amsfonts}
\usepackage{mathrsfs}
\usepackage{amsmath}
\usepackage{color}
\usepackage{graphicx}
\usepackage{bm}
\usepackage{amssymb}
\usepackage{xspace}
\usepackage{epstopdf}
\usepackage{dcolumn}
\usepackage{longtable}
\usepackage{multirow}
\usepackage{float}
\usepackage{comment}

\providecommand{\tabularnewline}{\\}

\usepackage[colorlinks=true, letterpaper=true, pdfstartview=FitV, linkcolor=blue, citecolor=blue, urlcolor=blue]{hyperref}

\makeatother

\begin{document}
\title{Low-Frequency Divergence of Circular Photomagnetic Effect in Topological Semimetals}

\author{Jin Cao}
\thanks{These authors contributed equally to this work}
\affiliation{Centre for Quantum Physics, Key Laboratory of Advanced Optoelectronic Quantum Architecture and Measurement (MOE), School of Physics, Beijing Institute of Technology, Beijing, 100081, China}
\affiliation{Beijing Key Lab of Nanophotonics \& Ultrafine Optoelectronic Systems, School of Physics, Beijing Institute of Technology, Beijing, 100081, China}

\author{Chuanchang Zeng}
\thanks{These authors contributed equally to this work}
\affiliation{Centre for Quantum Physics, Key Laboratory of Advanced Optoelectronic Quantum Architecture and Measurement (MOE), School of Physics, Beijing Institute of Technology, Beijing, 100081, China}
\affiliation{Beijing Key Lab of Nanophotonics \& Ultrafine Optoelectronic Systems, School of Physics, Beijing Institute of Technology, Beijing, 100081, China}

\author{Xiao-Ping Li}
\affiliation{Centre for Quantum Physics, Key Laboratory of Advanced Optoelectronic Quantum Architecture and Measurement (MOE), School of Physics, Beijing Institute of Technology, Beijing, 100081, China}
\affiliation{Beijing Key Lab of Nanophotonics \& Ultrafine Optoelectronic Systems, School of Physics, Beijing Institute of Technology, Beijing, 100081, China}

\author{Maoyuan Wang}
\affiliation{Centre for Quantum Physics, Key Laboratory of Advanced Optoelectronic Quantum Architecture and Measurement (MOE), School of Physics, Beijing Institute of Technology, Beijing, 100081, China}
\affiliation{Beijing Key Lab of Nanophotonics \& Ultrafine Optoelectronic Systems, School of Physics, Beijing Institute of Technology, Beijing, 100081, China}

\author{Shengyuan A. Yang}
\affiliation{Research Laboratory for Quantum Materials, Singapore University of Technology and Design, Singapore 487372, Singapore}

\author{Zhi-Ming Yu}
\email{zhiming\_yu@bit.edu.cn}
\affiliation{Centre for Quantum Physics, Key Laboratory of Advanced Optoelectronic Quantum Architecture and Measurement (MOE), School of Physics, Beijing Institute of Technology, Beijing, 100081, China}
\affiliation{Beijing Key Lab of Nanophotonics \& Ultrafine Optoelectronic Systems, School of Physics, Beijing Institute of Technology, Beijing, 100081, China}

\author{Yugui Yao}
\email{ygyao@bit.edu.cn}
\affiliation{Centre for Quantum Physics, Key Laboratory of Advanced Optoelectronic Quantum Architecture and Measurement (MOE), School of Physics, Beijing Institute of Technology, Beijing, 100081, China}
\affiliation{Beijing Key Lab of Nanophotonics \& Ultrafine Optoelectronic Systems, School of Physics, Beijing Institute of Technology, Beijing, 100081, China}

\begin{abstract}
Novel fermions with relativistic linear dispersion can emerge as low-energy excitations in topological semimetal materials. Here, we show that the orbital moment contribution in the circular photomagnetic effect for these topological semimetals exhibit an unconventional $\omega^{-1}$ frequency scaling, leading to significantly enhanced response in the low frequency window, which can be orders of magnitude larger than previous observations on conventional materials. Furthermore, the response tensor is directly connected to the Chern numbers of the emergent fermions, manifesting their topological character. Our work reveals a new signature of topological semimetals and suggests them as promising platforms for optoelectronics and spintronics applications.
\end{abstract}
\maketitle


Topological semimetals, which host unconventional emergent fermion modes around band nodal points at the Fermi level, have been a focus of research in recent years \citep{rmp2016_Das,rmp2016_Chiu,rmp2018_Ashvin_Weyl,np2015_Dai}. For example, in a Weyl semimetal, the electrons around the so-called Weyl nodal points acquire a relativistic linear dispersion and are described by the Weyl Hamiltonian with an emergent pseudospin-$1/2$ structure \citep{njp2007_Murakami,prb2011_wan_YIrO_weyl}. Recent studies showed that generalizations of Weyl fermions with higher pseudospins of the form $\sim \bm k\cdot \bm S$ also exist in crystalline materials and can be engineered in artificial structures \citep{science2016_BAB,EP_tyipeII,EP_TYPE_III,prb2021_Tang_band_nodes,prb2021_Tang_msg}. The unusual linear dispersion, the pseudospin structure, and the possible topological charge of these emergent fermions should exhibit unique signatures in many physical phenomena \citep{plb1983_ABJ_ChiralAnomaly,prb2013_Spivak_ChiralAnomaly,cpl2013_jhzhou_WSMs,prl2014_Burkov_ChiralAnomaly,npjqm2017_Guan_gravity,nm2019_review_Weng_weylopt,review2020_nagaosa,nm2020_Liu_review_weylopt,prb2021_Cano_multifold}, and this constitutes a main topic of current research on topological semimetals.

Circularly polarized light is a powerful probe of band topology \citep{prb2008_Yao_graphene,prl2012_xiao_tmd,prb2018_LiuYing}. As shown by de Juan \emph{et al.} \citep{nc2017_Juan_QCPGE}, it generates in Weyl semimetals a quantized injection current contribution to the circular photogalvanic effect, and the result can be extended to higher pseudospin cases \citep{prb2018_Flicker_QCPGE_multifold}. Since circularly polarized light carries an intrinsic angular momentum, its absorption in a material can generates a magnetization. This circular photomagnetic effect (CPME), also known as the inverse Faraday effect \citep{pr1963_Pershan_IFE,prl1965_IFE_exp,pr1966_Pershan_IFE,nature2005_Fe3_IFE,prl2009_IFE_Vahaplar,prb2014_Battiato_IFE,prb2016_IFE_Freimuth,prl2016_IFE_Berritta,npho2020_gold_IFE} or nonlinear Edelstein effect \citep{prb2021_Ju_NlEE}, has recently been considered in Weyl semimetals \citep{prb2018_IFE_DSMs,prb2020_IFE_graphene,arxiv2020_IFE_DSMs,arxiv2020_gao_IFE_spin,prl2021_IFE_DSMs}. Notably, Gao \emph{et al.} found that the spin contribution to CPME in Weyl semimetals is frequency independent ($\sim \omega^0$) and manifests the topological structure of Weyl fermions \citep{arxiv2020_gao_IFE_spin}.

In this work, we show that the orbital moment contribution to CPME, while vanishing for Weyl fermions, can generate a significant magnetization in topological semimetals with higher pseudospin fermions. Importantly, the result has a $\omega^{-1}$ scaling hence exhibits a low-frequency divergence, which makes it dominant at low frequencies. Moreover, we
reveal that the response tensor can be expressed in termed of the Chern numbers of bands involved in the optical transitions, exhibiting the topological character of the system. The corrections due to lattice effect and  the spin contribution are discussed and found to be subdominant in the low frequency window. Based on realistic parameters, the resulting magnetization can reach {$1\ \mu_B/\text{nm}^3$} under an infrared light with intensity of {${10}^{12}~\mathrm{W/m^2}$}, which can be readily detected in experiment. Our work discovers a nonlinear optical signature of topological semimetals and suggests these materials as promising platforms for opto-spintronics.

\begin{figure}
\begin{centering}
\includegraphics[width=7.0cm]{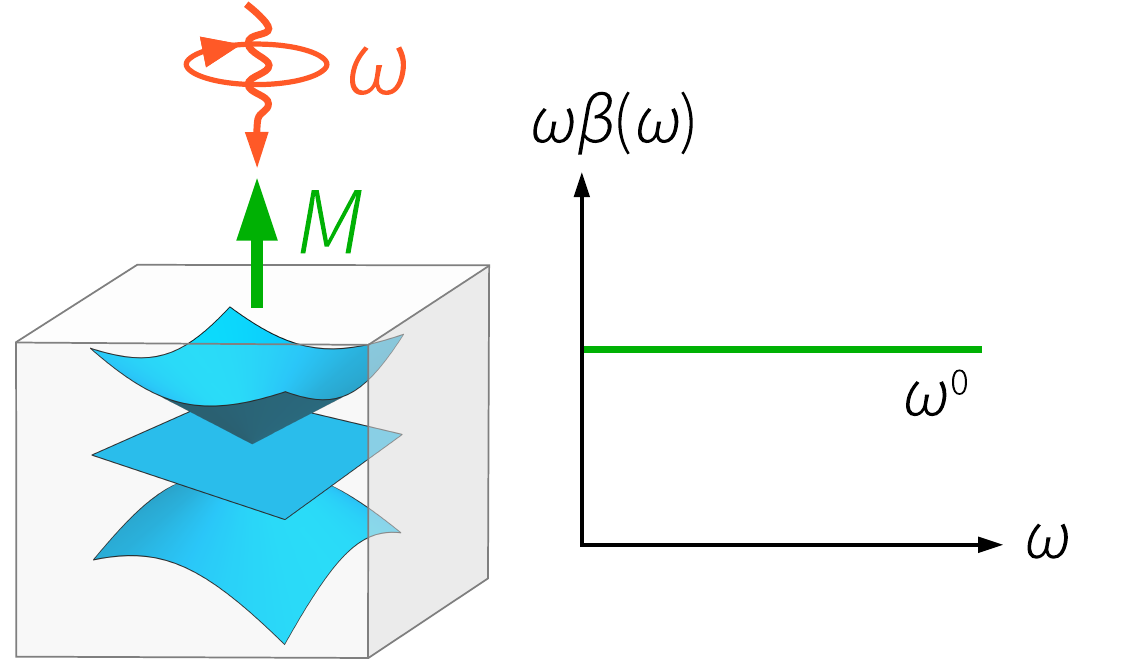}
\par\end{centering}
\caption{Sketch of CPME in topoligical semimetals with emergent relativistic fermion. Under a circular polarized light, a static magnetization can be generated. In this case, the product of frequency $\omega$ with response tensor $\beta(\omega)$ shows a flat plateau at low energies.}
\end{figure}

{\emph{\textcolor{blue}{General formula.}}}
We consider a three-dimensional (3D) \emph{nonmagnetic} solid under the irradiation of a circularly polarized light with frequency $\omega$. The induced magnetization corresponding to CPME should flip its sign under the reversal of circular polarization. It can be generally expressed as
{\begin{eqnarray}
M_{a} & = & \beta_{ab}\left(\omega\right)\left[i\boldsymbol{E}\left(\omega\right)\times\boldsymbol{E}\left(-\omega\right)\right]_{b},\label{eq:pheno-cpom}
\end{eqnarray}
where $\boldsymbol{E}$ is the electric field of light, $\beta$ is the CPME response tensor, and the subscripts $a$ and $b$ label the Cartesian components. The expression of $\beta$ can be derived by using the second-order perturbation theory. In general, there are both spin and orbital contributions to the induced magnetization. In this work, our focus is on the contribution by orbital magnetic moment, which within the single particle approximation can be put into the following form \citep{prb2021_Ju_NlEE,Supplemental_Materials} (we set $e=\hbar=1$)
\begin{eqnarray}
\beta_{ab}\left(\omega\right) & = & \frac{\pi \tau}{ V}\sum_{mn,\boldsymbol{k}}f_{nm,\boldsymbol{k}}\Delta_{mn,\boldsymbol{k}}^{a}
R_{mn,\boldsymbol{k}}^{b}\delta\left(\varepsilon_{mn,\boldsymbol{k}}-\omega\right).\nonumber \\
\label{eq:main}
\end{eqnarray}
Here, $\tau$ is the electron relaxation time, $V$ is the volume of the system, $f_{nm,\boldsymbol{k}}= f_{n\boldsymbol{k}}-f_{m\boldsymbol{k}}$ with $f_{n\bm k}$ the Fermi-Dirac distribution function of state $|u_{n\bm k}\rangle$, $R_{mn,\boldsymbol{k}}^{d}=i\epsilon^{dbc}r_{mn,\boldsymbol{k}}^{b}r_{nm,\boldsymbol{k}}^{c}$ with $\boldsymbol{r}_{mn,\boldsymbol{k}}=i\left\langle u_{m\boldsymbol{k}}|\partial_{\boldsymbol{k}}u_{n\boldsymbol{k}}\right\rangle$ ($m\neq n$) the interband Berry connection, $\varepsilon_{mn,\boldsymbol{k}}=\varepsilon_{m\boldsymbol{k}}-\varepsilon_{n\boldsymbol{k}}$ and $\Delta_{mn,\boldsymbol{k}}^{a}=\mu_{m\boldsymbol{k}}^{a}-\mu_{n\boldsymbol{k}}^{a}$ represent the  energy and  orbital magnetic moment differences between the states involved in the optical transition, where $\varepsilon_{m\boldsymbol{k}}$ is the band energy and $\mu_{m\boldsymbol{k}}^{a}$ is given by \citep{rmp2010_xiao_berry,prb1996_Chang,prl2005_Xiao_om,prl2007_shi_om,jpcm2008_Chang},
\begin{eqnarray}
\mu_{m\boldsymbol{k}}^{a} & = & -\epsilon^{abc}\frac{i}{2}\langle\partial_{k_b}u_{m\boldsymbol{k}}\big|[H(\boldsymbol{k})-\varepsilon_{m\boldsymbol{k}}]\big|
\partial_{k_c}u_{m\boldsymbol{k}}\rangle,
\label{eq:orbim}
\end{eqnarray}
with $H(\boldsymbol{k})$ the unperturbed Bloch Hamiltonian.

The formula (\ref{eq:main}) corresponds to a simple physical picture. Under the light irradiation, the optical transition between $|u_{n\bm k}\rangle$ and $|u_{m\bm k}\rangle$ would result in a change of orbital magnetic moment if $\Delta_{mn,\boldsymbol{k}}^a$ is nonzero. For a nonmagnetic system, the contributions from $\bm k$ and $-\bm k$ states would cancel each other if they have the same transition rates. The circular polarization of light breaks this symmetry and hence can result in a net magnetization. Furthermore, we note that the orbital moment $\mu^a_{n\bm k}$ is odd under time reversal and even under spatial inversion \citep{rmp2010_xiao_berry}. Therefore, the inversion symmetry must be broken for this contribution to be nonzero. {This means that the target system should belong to one of the 21 noncentrosymmetric point groups. They covers the gyrotropic and piezoelectric point groups, including $C_{1}$, $C_{2}$, $C_{s}$, $D_{2}$, $C_{2v}$, $C_{4}$, $S_{4}$, $D_{4}$, $C_{4v}$, $D_{2d}$, $C_{3}$, $D_{3}$, $C_{3v}$, $C_{6}$, $C_{3h}$, $D_{6}$, $C_{6v}$, $D_{3h}$, $T$, $O$, and $T_{d}$. }

In addition, $\Delta_{mn,\boldsymbol{k}}^a$ vanishes for a strict two-band system. Hence, the orbital moment contribution is expected to be small in systems with only two isolated bands in the optical transition range, such as the case for Weyl fermions. Nevertheless, as we will show below, the effect is significantly enhanced for topological semimetals with higher pseudospin fermions.

{\emph{\textcolor{blue}{Scaling relation for linear fermions.}}} Before doing any detailed calculations, we first argue that the response tensor in Eq.~(\ref{eq:main}) follows a $\omega^{-1}$ frequency scaling for higher pseudospin fermions due to their relativistic linear dispersion.

Consider a generic effective Hamiltonian for emergent fermions with linear dispersion:
\begin{eqnarray}
H & = & \boldsymbol{k}\cdot\boldsymbol{\Gamma},\label{eq:LHam}
\end{eqnarray}
where $\boldsymbol{k}$ is the momentum measured from the nodal point, and $\bm{\Gamma}$ is a vector of $k$-independent matrices with dimension corresponding to the degeneracy of the nodal point.
A crucial feature here  is that the eigenstate $|u_{n\boldsymbol{k}}\rangle$ only depends on direction of $\boldsymbol{k}$ vector, i.e., $\hat{\bm k}$, but not its magnitude $k$. 

Let's consider the following scaling transformation in momentum and frequency:
\begin{eqnarray}
 \boldsymbol{k}\rightarrow \boldsymbol{k}^{\prime}=\lambda \boldsymbol{k}, & \ \  &  \omega\rightarrow \omega^{\prime}=\lambda \omega ,\label{eq:resca1}
\end{eqnarray}
with $\lambda$ a real number.
For the linearly dispersing fermions in (\ref{eq:LHam}), one has  $H(\lambda \boldsymbol{k})  = \lambda  H(\boldsymbol{k})$ and $\varepsilon_{m,\lambda \boldsymbol{k}}=\lambda \varepsilon_{m,\boldsymbol{k}}$.
Moreover, due to the feature noted above, the eigenstates remain invariant under this rescaling, i.e., $ |u_{n,\lambda  \boldsymbol{k}}\rangle = |u_{n\boldsymbol{k}}\rangle$.
It follows that in Eq.~(\ref{eq:main}), $R_{mn,\lambda  \boldsymbol{k}}^{a}=\lambda^{-2}R_{mn,\boldsymbol{k}}^{a}$ and $\mu_{n,\lambda  \boldsymbol{k}}^{a}=\lambda^{-1}\mu_{n\boldsymbol{k}}^{a}$.
Consider the low-temperature regime, where $f_{n\boldsymbol{k}}\approx\Theta(E_F-\varepsilon_{n\boldsymbol{k}})$ with $\Theta$ the step function and $E_F$ the Fermi energy.
Then the 3D integral in Eq.~(\ref{eq:main}) will be  reduced to a 2D integral performed over the optical transition surface $S_{OT}$ consisting all the points in the momentum space where the optical transition occurs, i.e., where the quantity
$f_{mn,\bm k}\delta(\varepsilon_{mn,\bm k}-\omega)$ is nonzero for some $m$ and $n$. Then one can show that as long as $S$ remains a closed surface enclosing the nodal point under rescaling, the orbital moment CPME tensor follows the simple scaling relation:
\begin{eqnarray}\label{scale}
 \beta_{ab}(\lambda \omega)=\lambda^{-1} \beta_{ab}(\omega).
\end{eqnarray}
In other words, $\beta_{ab}$ scales as $\omega^{-1}$. If one plots the quantity $\omega\beta_{ab}$ versus frequency, it should exhibit a flat plateau for linear fermions.

{
\global\long\def\arraystretch{1.4}%
\begin{table*}
\caption{\label{Tab} 
Space groups that can host the linearly-dispersing emergent fermions. 
Here C-2 TP and C-4 DP stand for the charge-2 triple point and charge-4 Dirac point, respectively. 
Spin-1 (Spin-3/2) particle is a special case of C-2 TP (C-4 DP).
The form of the effective Hamiltonian ${\cal H}_{\text{eff}}$ are explicitly presented in SM~\citep{Supplemental_Materials}. The column with “with SOC” indicates whether the system contains spin-orbit coupling.}
\begin{ruledtabular}
\begin{tabular}{llll}
Notation  & SG and Location  & ${\cal H}_{\text{eff}}$ & with SOC \tabularnewline
\hline
C-2 TP (spin-1 particle) &  195, $\Gamma$, R; 196, $\Gamma$; 197, $\Gamma$, H; 198, $\Gamma$;
                                199, $\Gamma$, H; 207, $\Gamma$, R;  & $v_{1}\boldsymbol{k}\cdot\boldsymbol{S}$ & N  \tabularnewline
 &   208, $\Gamma$, R; 209, $\Gamma$; 210, $\Gamma$;  211, $\Gamma$, H; 212, $\Gamma$; 213, $\Gamma$; 214, $\Gamma$, H; &  &  \tabularnewline
 \hline
C-2 TP &  197, P; 211, P  & $v_{1}\boldsymbol{k}\cdot\boldsymbol{S}+v_{2}\boldsymbol{k}\cdot\boldsymbol{S}^{\prime}$  & N \tabularnewline
C-2 TP &   199, P; 214, P  & $v_{1}\boldsymbol{k}\cdot\boldsymbol{S}+v_{2}\boldsymbol{k}\cdot\boldsymbol{S}^{\prime}$  & Y \tabularnewline
\hline
C-4 DP & Same as spin-1 particle & $v_{1}\boldsymbol{k}\cdot\boldsymbol{S}+v_{2}\boldsymbol{k}\cdot\boldsymbol{S}^{\prime}$ & Y \tabularnewline
\end{tabular}\end{ruledtabular}
\end{table*}

}

There are two remarks about the scaling relation (\ref{scale}). First, the analysis shows that the relation is solely due to the linear dispersion of the emergent fermion.
The result is general in that it does not depend on the specific form of the model (the $\Gamma$ matrices), the degree of degeneracy of the nodal point, nor the possible anisotropy in the dispersion. Second, the scaling shows that for nodal points sitting at the Fermi level, the CPME can be divergently large at low frequencies. (Obviously, when Fermi energy deviates from the nodal point, the effect has a lower cutoff frequency due to Pauli blocking.) Since spin and other contributions at most have a $\omega^0$ scaling at low frequencies, the orbital moment contribution should dominate in this regime \citep{arxiv2020_gao_IFE_spin}. In the following, we perform concrete model calculations to illustrate these points.

{\emph{\textcolor{blue}{Result for pseudospin-$j$ fermions.}}}
For a concrete study, we take pseudospin-$j$ fermions described by the Hamiltonion \citep{science2016_BAB,EP_tyipeII,prb2021_yang_TP,Supplemental_Materials}
\begin{eqnarray}
H(\boldsymbol{k}) & = & \chi v_{F}\boldsymbol{k}\cdot\boldsymbol{S}.\label{eq:spinj}
\end{eqnarray}
As mentioned above, the anisotropy does not affect the scaling relation, so here we just take an isotropic Fermi velocity $v_F$. $\chi=\pm 1$ represents a kind of chirality, and $\bm S$ is a vector of the spin-$j$ matrices, satisfying $\left[S_{a},S_{b}\right]=i\epsilon^{abc}S_{c}$.

We will show that the $\beta_{ab}$ is closely connected to the topological charge (the Chern number) of the fermions. For the $n$-th band (counted from bottom to top) of the pseudospin-$j$ particle, the Chern number is given by
\begin{equation}
  {\cal{C}}_n=-2\chi(j+1-n).
\end{equation}
Here, the Chern number is defined on a closed surface surrounding the nodal point.
The  orbital moment for the $n$-th band is obtained as
$\boldsymbol{\mu}_{n\boldsymbol{k}} = e\chi v_{F}\frac{\boldsymbol{\hat{k}}}{8k}\left[4j\left(j+1\right)-{\cal{C}}_n^{2}\right]
$.
Interestingly, $\boldsymbol{\mu}_{n\boldsymbol{k}}$ depends on the absolute value of ${\cal{C}}_n$ but not its sign. Hence, bands with opposite Chern numbers would have the same orbital moment.

\begin{figure}
\begin{centering}
\includegraphics[width=8.6cm]{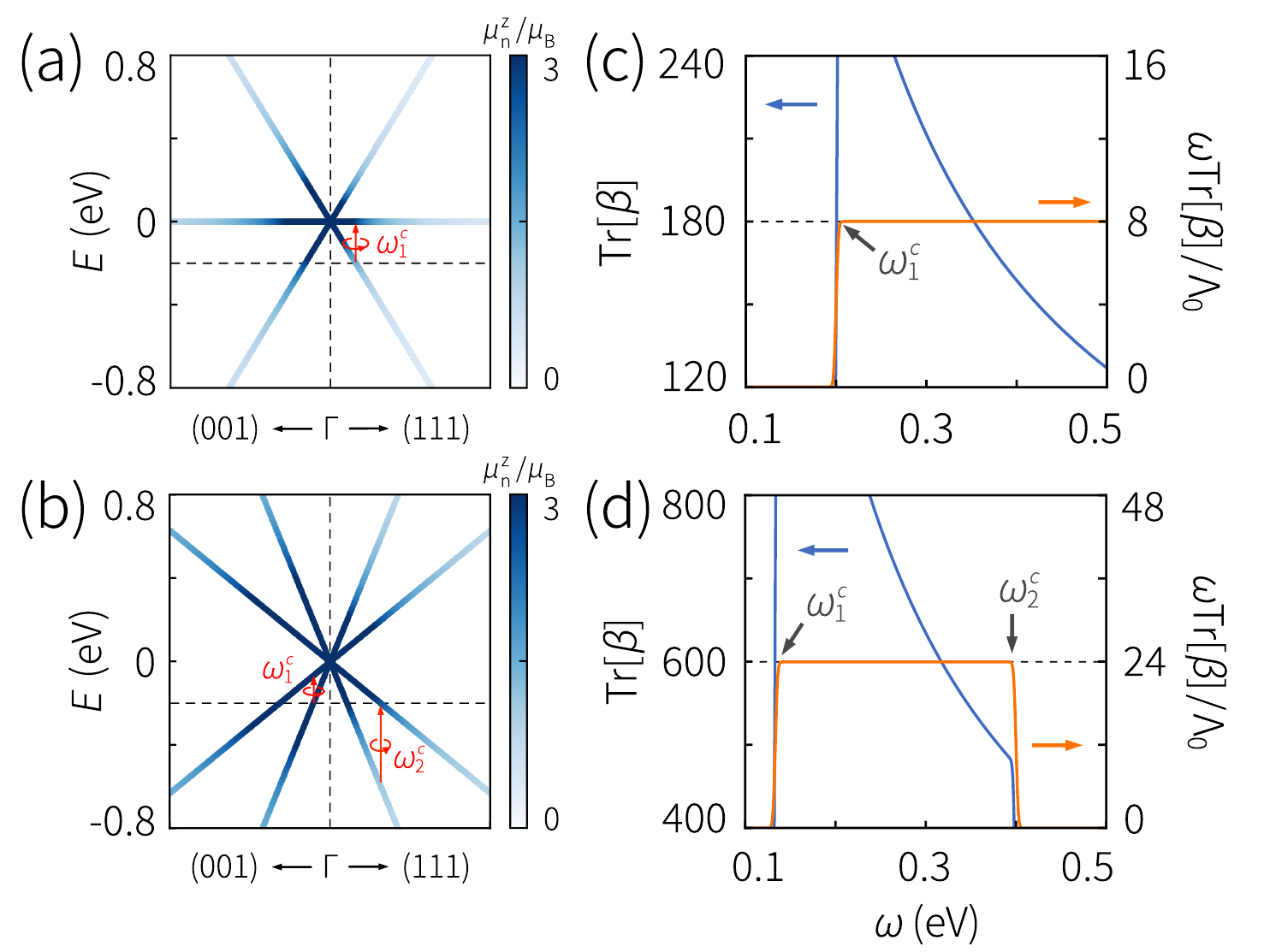}
\par\end{centering}
\caption{CPME for pseudospin-$j$ fermions. (a) The band structure and (c) the CPME at a fixed Fermi energy [horizontal dashed line in panel (a)] for spin-$1$ fermions. Similar results are also shown in panel (b) and (d) for spin-$3/2$ fermions.
The color map in (a) and (b) indicates the orbital moment of the states.
The frequencies where the CPME plateau starts and ends are marked by the black arrows. Here we set $E_F=-0.2$~eV, $\chi=1$, $v_F=2~\mathrm{eV\cdot\mathring{A}}$ and $\tau=1$~ps, and $\mathrm{Tr}[\beta]$ is in the unit of $\mu_{B}/(\mathrm{V^{2}\cdot\mathring{A}})$ with  $\mu_B$ the  Bohr magneton. \label{Fig_2} }
\end{figure}

The pseudospin-1/2 case corresponds to the Weyl fermions, which, as we discussed before, have a vanishing orbital moment contribution. In the following, we will focus on pseudospin-1 and 3/2 fermions. For $j>3/2$, the qualitative feature of the result remains but the expression gets more complicated, and such nodal points are not directly stabilized by crystalline symmetry \citep{EP_tyipeII}.

Considering optical transitions for pseudospin-1 and 3/2 fermions, we find that
for both cases, the transition is allowed only between the nearest two bands~\citep{Supplemental_Materials}.
Moreover, let's focus on the trace of the  CPME tensor in (\ref{eq:main}), because it can be put into a compact form
\begin{eqnarray}
\mathrm{Tr}\left[\beta\left(\omega\right)\right] & = & \frac{\tau}{8\pi^2 }\oint_{S_{OT}}d\tilde{\mathcal{S}}_{mn,\boldsymbol{k}}\boldsymbol{\Delta}_{mn,\boldsymbol{k}}\cdot\boldsymbol{R}_{mn,\boldsymbol{k}},
\label{eq:beta_simply}
\end{eqnarray}
where $m=n+1$, $d\tilde{\mathcal{S}}_{mn,\boldsymbol{k}}=d\mathcal{S}_{mn,\boldsymbol{k}}/\left|\partial_{\boldsymbol{k}}\varepsilon_{mn,\boldsymbol{k}}\right|$
is a reduced area element over the optical transition surface.
Meanwhile, the off-diagonal elements of $\beta\left(\omega\right)$ vanish for the isotropic model.

By straightforward calculation, we find that the results for spin-1 and 3/2 fermions can be put into the following unified form. For $E_F<0$, we have
\begin{eqnarray}
\mathrm{Tr}\left[\beta\right]=\frac{\Lambda_{0}}{\omega}\left|\mathcal{C}_{1}\right|\left(\mathcal{C}_{1}^{2}-\mathcal{C}_{2}^{2}\right),\label{eq:qcpom1}
\end{eqnarray}
and for  $E_F>0$,
\begin{eqnarray}
\mathrm{Tr}\left[\beta\right]=-\frac{\Lambda_{0}}{\omega}\left|\mathcal{C}_{2j+1}\right|\left(\mathcal{C}_{2j+1}^{2}-\mathcal{C}_{2j}^{2}\right),\label{eq:qcpom2}
\end{eqnarray}
where $\Lambda_0= \tau v_{F}/(32\pi)$. This result explicitly demonstrates the $\omega^{-1}$ scaling of the $\beta$ tensor. Although the result is given for the trace, it is clear that the scaling holds for each tensor element. In Fig.~\ref{Fig_2}, we plot the result from numerical calculation. The clear exhibition of a flat plateau in $\omega \mathrm{Tr}\left[\beta\right]$ confirms the scaling behavior. In the figure, one also observes the lower cutoff $\omega^c_1$ of the divergence when the Fermi energy deviates from the nodal point. For the pseudospin-1 (3/2) model, we have $\omega^{c}_1=|E_F|$ ($=2|E_F|/3$). For the pseudospin-3/2 model, there is also an upper cutoff at $\omega^{c}_2=2|E_F|$.

Besides the $\omega^{-1}$ scaling, a salient feature in the result Eq.~(\ref{eq:qcpom1}-\ref{eq:qcpom2}) is that the orbital moment CPME is closely connected to the topological charges. Furthermore, only the absolute value of the charge comes in, not its sign. Thus, for a pair of such nodal points with opposite chirality $\chi=\pm$, the net result should double rather than canceling each other. This is distinct from the injection current contribution to the circular photogalvanic effect, where points with opposite chirality would cancel the net effect \citep{nc2017_Juan_QCPGE,prb2018_Flicker_QCPGE_multifold,SciAdv2020_exps_RhSi_QCPGE}, so that effect can only happen for chiral point groups
which eliminates all possible mirrors. In this sense, the constraint for the effect here is less stringent.

To guide the realization of spin-$1$ and $3/2$ fermions,
we list the space groups that can stabilize these fermions in 3D nonmagnetic crystals. Meanwhile, such fermions may also appear without symmetry protection at a critical state. In that case, one would expect a significant enhancement of the CPME response when the critical state is approached.

\begin{figure}
\begin{centering}
\includegraphics[width=8.6cm]{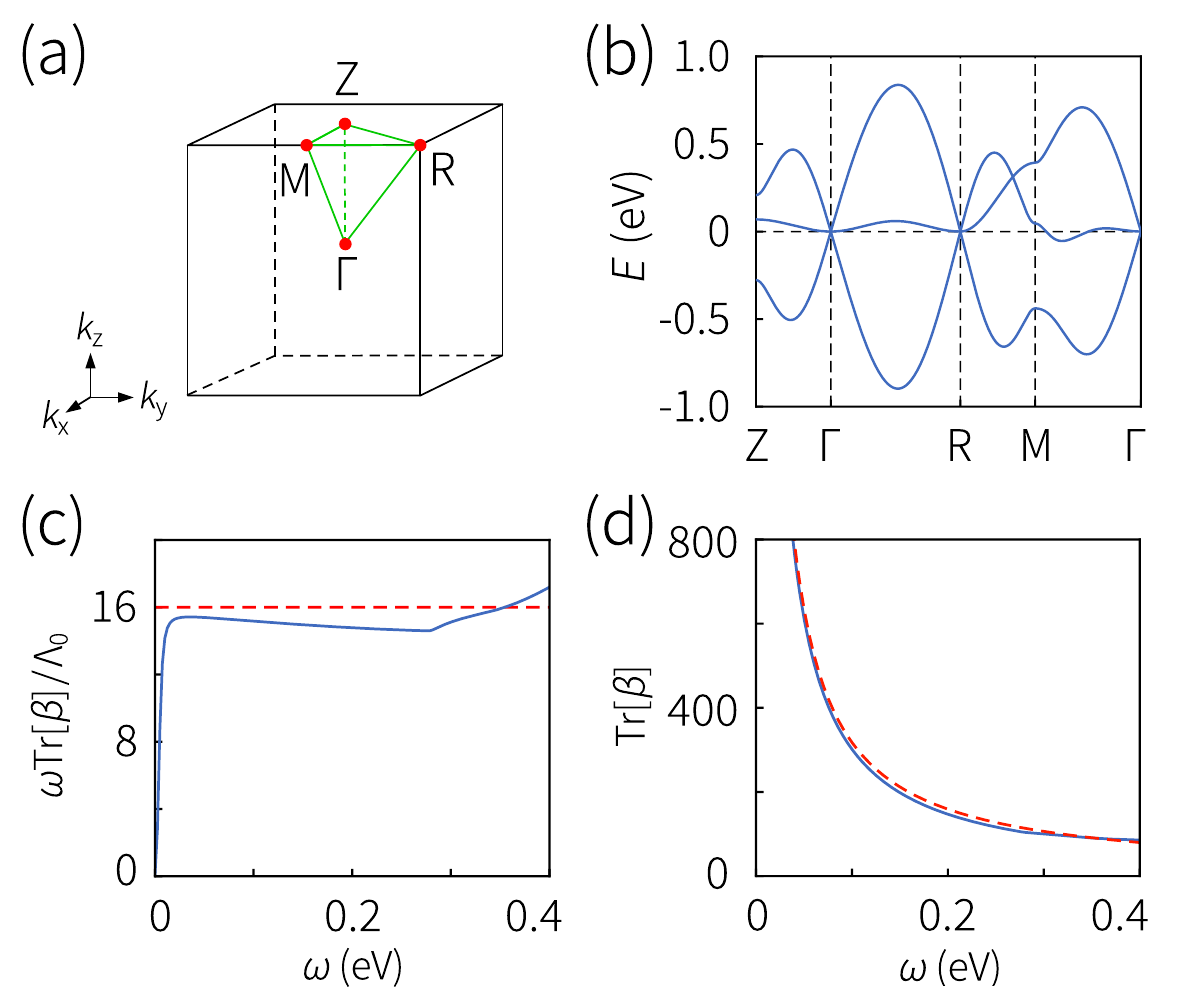}
\par\end{centering}
\caption{Result from lattice model calculation. (a) Brillouin zone of SG 195. (b) The band structure of a lattice model with spin-1 fermions.
(c-d) Numerical result (in blue) of the response tensor from the lattice model. The red dash line denotes the analytical result from the effective $k\cdot p$ model. We set $E_F=0$ eV here and the parameters of the lattice model are presented in SM~\citep{Supplemental_Materials}. \label{Fig_3}
}
\end{figure}

{\emph{\textcolor{blue}{Lattice model calculation.}}}
Our analysis so far is based on the effective model around a nodal point in a topological semimetal. In a real system, the effective model is valid in a limited range in the energy momentum space. The lattice effect will introduce higher order (in $k$) corrections to the effective model.

For example, let's consider adding a quadratic correction term $H^\prime\sim k^2$ to the effective Hamiltonian Eq.~(\ref{eq:LHam}).
In the low frequency regime, this term can be treated as a perturbation.
The original eigenstate $|u_0\rangle$ would be perturbed into
$|u\rangle=|u_0\rangle+ |u^{\prime}\rangle$ with $|u^{\prime}\rangle \sim k$. Then
the correction for $\beta$ scales as $\beta_{ab}^{\prime} \propto \omega^{0}$ for small $\omega$.
Therefore, similar to the spin contribution, the correction from lattice effect is subdominant in the low frequency regime.

To confirm this point, we consider a lattice model of pseudospin-$1$ fermions and numerically assess the correction from lattice effect. The model belongs to the space group No.~195 and is presented in \citep{Supplemental_Materials}. Its band structure is plotted in Fig.~\ref{Fig_3}(b).
There are two pseudospin-1 nodal points at $\Gamma$ and R points in the Brillouin zone (see Table \ref{Tab}).
The two points have opposite chirality.
The Chern  number for the point at $\Gamma$ (R) point is calculated as ${\cal{C}}=2$ (${\cal{C}}=-2$) for the lowest band.
From our result in Eq.~(\ref{eq:qcpom1}),
one expects that the contributions from the two points would add up and a plateau should appear in $\omega \text{Tr}[\beta]$ at low frequencies. This is indeed verified in Fig.~\ref{Fig_3}(c), where for reference we also plot the result from the effective linear model (the red dashed line).

{\emph{\textcolor{blue}{Discussions.}}}
Experimentally, the large CPME predicted here can be {probed by the magneto-optical Kerr microscopy}~\citep{nature2017_kerr1,nature2017_kerr2}.
As an estimation, the induced magnetization can be related to the light intensity as $M=GI$, with $I$ the
intensity of the incident light.
At time scale $t \gg \tau$, the photoinduced $M$ saturates to a constant value.
For a topological semimetal with spin-$1$ fermions at the Fermi level, we have $G=\frac{16 \Lambda_0}{\varepsilon_{0}c\omega}$, where $\varepsilon_0$ is the vacuum permittivity and $c$ is the speed of light.
Taking typical values of  $\tau= 1 $ ps, $v_F=4\times 10^5 $ m/s (e.g., as in the topological semimetal RhSi \citep{prb2017_Tangpeizhe,prl2017_Changguoqing,nature2019_Sanchez_RhSi,SciAdv2020_exps_RhSi_QCPGE}), and an infrared pump light with $\omega=0.5$~eV, $G$ is estimated as $G\sim 1.3\times10^{5}~\mu_{B}\mathrm{\mathring{A}}^{-1}/\mathrm{W}$. Under a moderate light intensity of {${10}^{12}~\mathrm{W/m^2}$}, the induced magnetization can reach {$\sim 1\ \mu_B/\text{nm}^3$}.
This large induced magnetization is at least two orders of magnitude larger than that observed in DyFeO$_3$ \citep{nature2005_Fe3_IFE} or gold nanoparticles \citep{npho2020_gold_IFE}, hence it is readily detectable in experiment.



%

\end{document}